\documentclass{elsart}

\journal{Chemical Physics Letters}
\usepackage{graphics}
\usepackage{epsfig}

\usepackage{amssymb}

\def\BNF{B$_{24}$N$_{24}$ }
\def\BNE{B$_{28}$N$_{28}$ }
\def\BNT{B$_{32}$N$_{32}$ }
\def\BNS{B$_{36}$N$_{36}$ }
\def\BNR{B$_{96}$N$_{96}$ }

\begin{document}

\begin{frontmatter}



\title{ Are hemispherical caps of boron-nitride nanotubes possible?}
\author[GW]{Rajendra R. Zope \corauthref{ZCOR}}
\ead{rzope@alchemy.nrl.navy.mil}
\author[NRL]{Brett  I. Dunlap}
\ead{dunlap@nrl.navy.mil}
\address[GW]{Department of Chemistry, George Washington University, Washington DC, 20052}
\corauth[ZCOR]{Fax: +1-202-767-1716}

\address[NRL]{Code 6189, Theoretical Chemistry Section, US Naval Research Laboratory,
Washington, DC 20375}

\date{\today}

\begin{abstract}
   We report all-electron, density-functional calculations with large Gaussian
polarization basis set of the recently synthesized 
octahedral $O$  \BNF cage that is perfectly round by symmetry, and boron-nitride (BN) 
clusters that its existence might suggest.  We consider whether it is energetically 
possible that the two halves of this round cage could cap the BN nanotubes, modeled 
by \BNE and \BNT. The energetics show that BN nanotubes with such round caps, are only 
slightly less favorable than the BN clusters containing six squares as the only
defects in the otherwise perfect hexagonal lattice.  A larger \BNR $O$ cage formed from
\BNF by adding sufficient hexagons to isolate all squares is not very favorable energetically.
\end{abstract}


\end{frontmatter}

	Boron-nitride clusters have been made by vacuum arc melting, dissolved in pyridine, and 
detected using time-of-flight mass spectroscopy, which shows that the \BNF cluster is made in
abundance \cite{Oku03}.  Its structure is likely to be the largest, and thus most round, alternate
boron-nitride (BN) cage in which all atom pairs are equivalent by
octahedral symmetry, in analogy to the icosahedral C$_{60}$ cluster. The latter cluster is a fullerene,
which is a hexagonal graphite sheet made to close into a cage by twelve pentagons as required 
by Euler's theorem \cite{KHOCS85}, for which there are no alternate BN cages \cite{ZSK97}.
Alternate BN cages are made by substituting alternate $sp^2$ carbon atoms of certain fullerene-like
cages with boron and nitrogen atoms. These cages contain only squares, hexagons, octagons, etc., and
only strong BN nearest-neighbor bonds.
The simplest and most stable alternate BN cages contain hexagons that are closed by exactly six
isolated squares \cite{SSL95}. However, \BNF is the alternate cage for the perfectly round 
C$_{48}$ cluster with $O_h$ symmetry \cite{Haymet85}.  Both are made round by the addition of twelve
defects--six octagons and the corresponding six squares required by Euler's theorem. When made from carbon this
cluster is quite unfavorable \cite{DT94}, but apparently not when made from boron nitride.
Our calculations show that the \BNF cage containing six squares isolated by hexagons is 
about 4-5 eV lower than the octahedral \BNF cage\cite{SSL95}.
Perhaps BN cages are almost as versatile as fullerene cages; that could be
the case if defects in boron nitride
cages cost significantly less energy than they do in fullerene cages.

    A study of the smaller fullerenes suggests that it is 
never favorable to replace, in the context of Euler's theorem, two pentagons with a 
single square \cite{Fowler96}.  So in carbon cages squares are likely to be twice as energetically 
unfavorable as pentagons, which cost, when isolated, at least 2.3 eV each \cite{Dunlap97}.  Heptagons and octagon
are less energetically favorable in carbon cages, but are required to bend and twist free-standing 
nanotubes \cite{Dunlap97}.  The energetics of fullerenes if applied to \BNF would argue strongly against
its existence.

    The most abundant fullerenes are buckminsterfullerene, C$_{60}$, and C$_{70}$.  The latter 
can be viewed as being formed by cutting buckminsterfullerene into hemispheres in a highest-symmetry 
direction and inserting a ring of ten atoms, and then rotating the top hemisphere half a revolution
before 
reconnecting it.  It is natural to think about extending that process to larger capsules \cite{HBR86}
and ultimately to the (5,5) carbon nanotube in the infinite limit \cite{MDW92,Ijima91}. 
Contrary to early belief, C$_{60}$ is not the most stable fullerene. The larger icosahedral fullerenes
are more stable \cite{DBMMW91}, but C$_{60}$  is the most stable in a wide range of cluster sizes and 
shapes \cite{SSKH98}.  Perhaps perfect roundness facilitates making C$_{60}$ and \BNF$\!$.

          If, on the other hand, extra defects in boron nitride fullerenes are so costly
that they never occur except in \BNF$\!\!$, then boron nitride caps have a very special shape.
If half a defect-free BN fullerene caps a BN nanotube just as half a defect-free fullerene caps a 
carbon nanotube \cite{Dresselhaus92} then each BN cap has precisely three squares.
Furthermore, the three squares must lie on the periphery of the nanotube, and, as three points determine 
a plane, the BN nanotube cap must be flat.  Defect-free flat and triangular BN caps are quite different
from hemispherical carbon nanotube caps.
	
      If \BNF is spherical and under some circumstances abundant, it is natural to 
ask if an additional ring of BN atoms can be added to form a less abundant but still 
possible cluster that is reminiscent of C$_{70}$, and if further rings are possible to 
make hemispherically capped BN nanotubes.  We study the energetics of this possibility using
analytic density functional theory.
	
                 Our calculations use the Slater-Roothaan method.  It uses Gaussian bases to 
fit the orbitals and Kohn-Sham potential \cite{KS65} of density-functional theory. The method
through robust 
and variational fitting is analytic and variational in all basis sets \cite{Dunlap03}.  
The most general functionals that it can treat so far are certain variants of the X$\alpha$
functional \cite{Vauthier}.
 In particular, it can handle different {$\alpha$'s}  on different elements analytically 
and variationally so that the atomized energies of any cluster can be recovered exactly,
 and all energies are accurate through first order in any changes to any basis set.  
Thus the energy does not change if we choose to constrain any fit, which we  do 
for all fits at no cost to the calculated energetics and negligible cost in computer time. 
Our work has spawned research efforts that brought both useful fitting techniques and 
density-functional theory to Gaussian-based {\sl ab initio} quantum chemistry.  Thus there are 
good basis sets for fitting the Kohn-Sham potential.  The best {\sl s-}type fitting bases are 
still those scaled from the {\sl s-}part of the orbital basis \cite{Dunlap79}. A package of basis sets has 
been optimized \cite{GSAW92} for use with DGauss \cite{AW91}.  We use the valence double-$\zeta$ orbital basis set 
DZVP2 and the {\sl pd} part of the (4,3;4,3) (A2) charge density fitting basis. 
Ahlrichs' group has generated a RI-J basis for 
fitting the charge density of a valence triple-$\zeta$ orbital basis set used in the Turbomole 
program \cite{EWTR97}.  We use its {\sl pdf} part with the standard 6311G** basis set downloaded from 
PNNL (http://www.emsl.pnl.gov:2080/forms/basisform.html).  For both sets of calculations the values 
of $\alpha$ chosen for B and N are,
 respectively, 0.786751 and 0.767473, which give the exact atomic energies \cite{clemmenti}
when used with the triple-$\zeta$/RI-J basis.  The accuracy of this method using Hartree-Fock 
values of $\alpha$ \cite{Schwarz72} away from the sides of the periodic table is roughly that of Becke's
half-and-half method \cite{Becke93}, but as it requires no numerical integration.  The method is 
ideal for use with Gaussian basis sets \cite{Dunlap03}.  Forces are computed nonrecursively using 
the 4-j generalized Gaunt coefficients \cite{Dunlap02}.

\begin{figure}
\epsfig{file=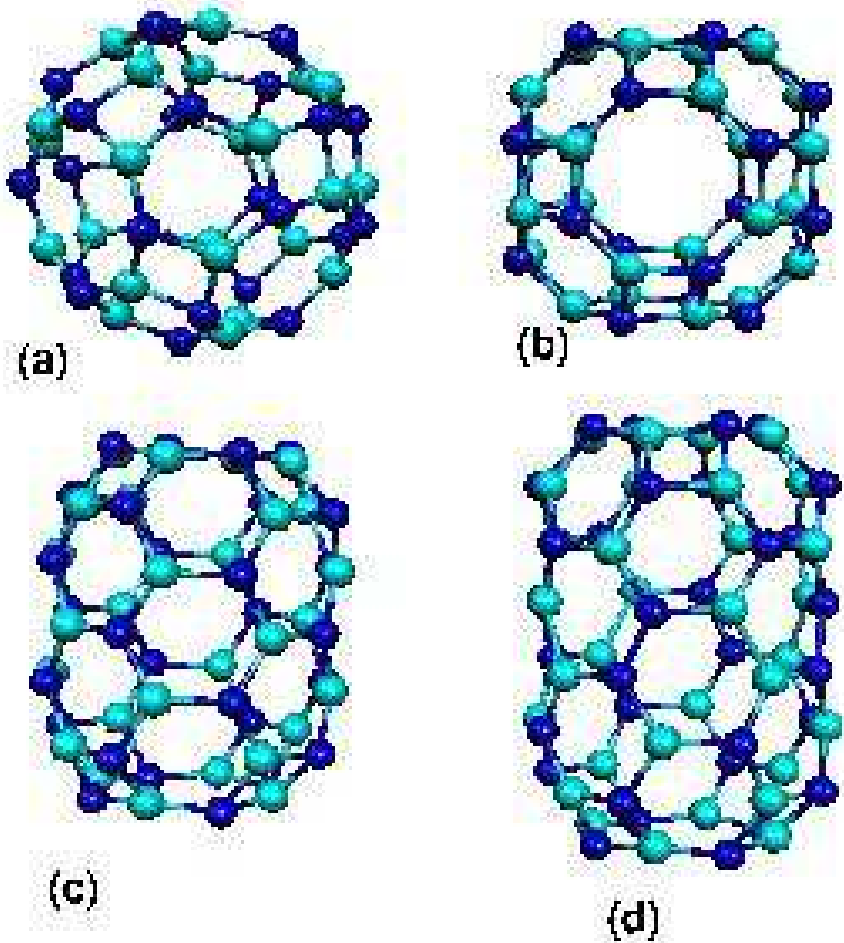,width=\linewidth,clip=true}
\caption{\label{fig1} (Color online) The optimized structures for hemispherically-capped BN nanotubes. (a) and (b) are
two different views of the \BNF cage: (a) along the C$_3$ axis,  (b) along C$_{4}$ axis.  (c) \BNE ({\sl C$_{4h}$}) cage 
obtained by adding a ring of eight alternating B and N atoms, (d) tubular \BNT ({\sl S$_8$}) cage by inserting
two rings of eight alternating B and N atoms (see text for more details). }
\end{figure}

\begin{table}
\caption{The binding energy per BN pair (BE), the energy gap between the highest 
occupied molecular orbital and the lowest unoccupied molecular orbital, and the vertical ionization
potential (VIP) for the optimized BN cages, all in eV.}
\label{table:BE}
\begin{tabular}{llrlrl}
\hline
             & Symmetry  &\multicolumn{3}{c}{6311G**/RI-J} & {DZVP2/A2}   \\
             &           &  BE     &    GAP        & VIP       &   BE   \\
\hline         
\BNF        &  \sl{O}      &   14.63       &  4.94   & 6.66   &  14.64 \\
\BNE        & \sl{C$_{4h}$}  &   14.89       &  4.86   & 8.48  &  14.90 \\
\BNE        & \sl{T}      &   14.95       &  5.38   & 8.64  &  14.95 \\
\BNT        & \sl {S$_8$}    &   14.96       &  4.95   & 8.36  &  14.98 \\
\BNS        & \sl {T$_d$}    &   15.06       &  5.33   & 8.42  &  15.11 \\
\BNR        & \sl {O}    &   14.97       &  4.25   &  7.74 &  15.12 \\
\hline
\end{tabular}
\end{table}

                The symmetry restricted optimization of the BN cages was performed using the 
BFGS computer code \cite{BFGS}. The optimized hemispherically capped BN nanotubes are shown in Fig. ~\ref{fig1}
while their symmetries are given in the Table I.  The (4,4) nanotubes are generated by extending along
the four-fold axis [Fig ~\ref{fig1}(b)].  The caps start off perfectly round by the symmetry of \BNF.
They remain quite round after geometry optimization. In the figure they look a little flat at the
tops and bottoms because the octagons take up a significant portion of their caps.  The (3,3) BN
nanotube generated by extending the three-fold direction would have rounder caps, but are probably less
likely to be made because they contain significantly more curvature.  The binding energy per BN pair (BE), 
the energy gap between the highest occupied molecular orbital (HOMO) and the lowest unoccupied 
molecular orbital (LUMO), and the vertical ionization potential calculated at the optimized 
geometries are also given in Table I.

\begin{figure}
\epsfig{file=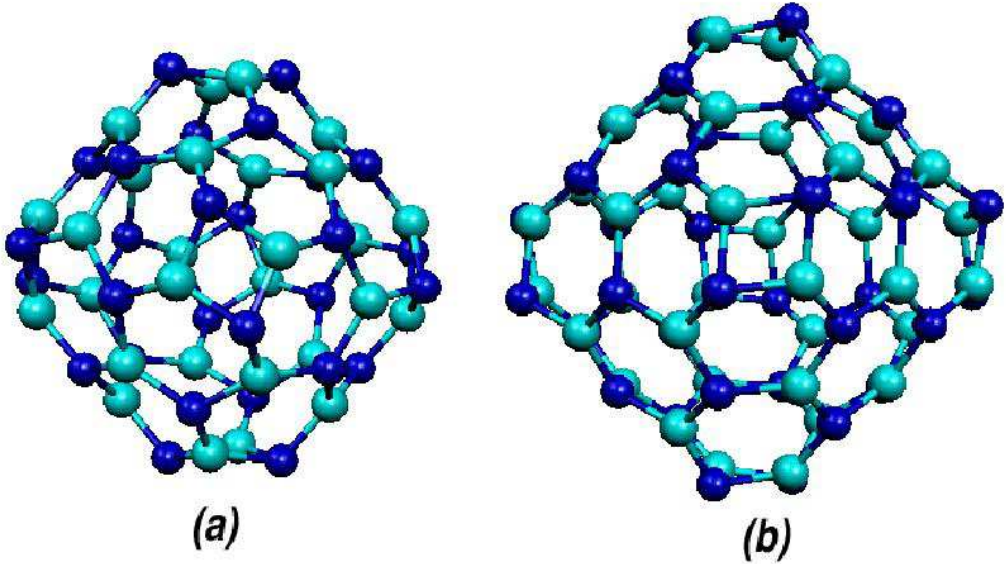,width=\linewidth,clip=true}
\caption{\label{fig2} The optimized structures for the favorable \BNE (left) and \BNS (right) cages.}
\end{figure}

The coordinates (in \AA) of the B and N atoms optimized
with 6311G**/RI-J basis set for the recently synthesized octahedral  \BNF  cage  
are (0.66, 1.66, 2.59) and (2.74, 1.65,  0.74) respectively for the three-fold axis in the 
(1,1,1) direction.  The  positions of  all remaining 
of atoms in the cage  can  be  generated by symmetry operations.  The HOMO and LUMO both
have $e$ symmetry and  yield a large  gap of 4.94 eV  that  agrees with the value of 4.95 eV
\cite{Oku03} obtained using DV-X$\alpha$ at the PM5 geometry.

                    The C$_{4h}$ \BNE cage [Fig.~\ref{fig1}(c)] is generated from the base \BNF 
cage by cutting the latter into two halves after orienting it along the C$_4$ axis,
then inserting a ring of eight alternating B and N atoms perpendicular to the axis,
i.e. horizontally, and then rotating the top half by an eighth of a revolution.
The resultant cage structure has {\sl C$_{4h}$} symmetry with eight inequivalent
atoms.  If another ring of four BN dimers is inserted in the same manner, then the
resultant \BNT cage is a tubular structure with {\sl S$_8$} symmetry [Fig.~\ref{fig1}(d)].
Table I shows systematic increase in the binding
energy by going from \BNF to the \BNE cage (0.26 eV per BN pair) and from \BNE to \BNT cage
(0.06 eV/BN pair).  The successive additions of more BN rings will result in  further 
stabilization of the BN tubular cages and will ultimately lead to the  (4,4) BN nanotube.
Thus our total energy calculations do not rule out the possibility of BN 
tubes with round caps similar to those observed in case of carbon nanotubes.
A recent molecular dynamics study of the growth mechanism
of BN nanotubes has also noted the formation of a cap containing four squares and octagon
for this nanotube \cite{XVCC98}. Molecular models of caps with four squares, octagons and hexagons
were also speculated earlier\cite{Kroto}.
All the BN cages studied  have rather  large HOMO-LUMO gaps of about 4.9-5.5 eV  similar to those observed for
the infinite BN nanotubes ($\sim 5.5$ eV)\cite{XRLC94}.  
The vertical ionization potential (VIP) 
calculated as the energy difference of the two self-consistent calculations for 
the neutral and singly positively  charged cage also show relatively large values. 

\begin{figure}
\epsfig{file=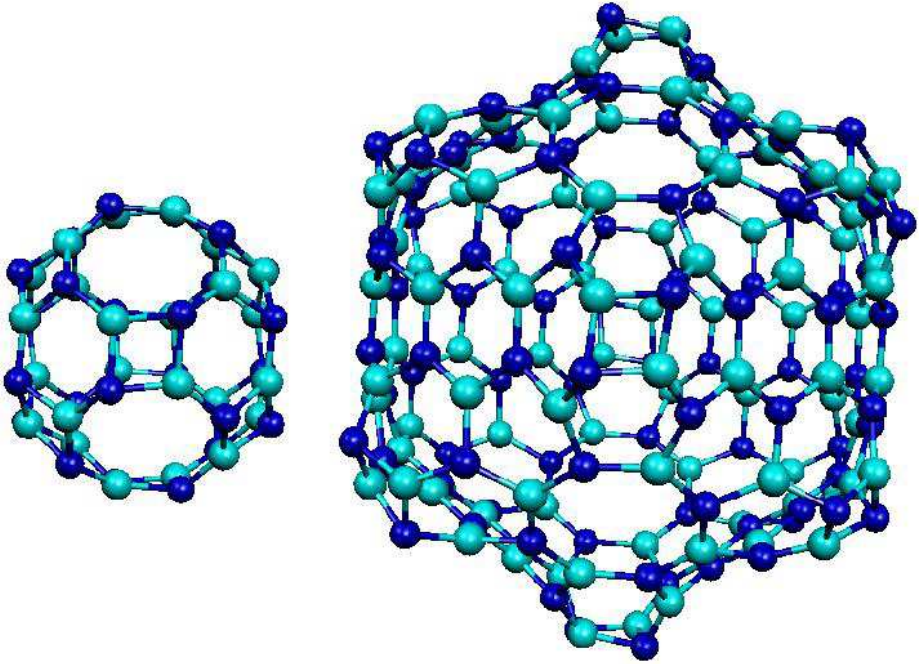,width=\linewidth,clip=true}
\caption{\label{fig3} (Color online) The optimized structures for the (a) \BNF  and  (b) \BNR  cages.}
\end{figure}

Fig. \ref{fig2} shows two examples of the most favorable BN cages that are composed of 
eight equivalent triangular faces meeting at six isolated squares \cite{SSL95,ZSK97}.
Their BE, HOMO-LUMO gap and VIP are given in Table \ref{table:BE}. 
The smaller cage, $T$ \BNE,
Fig. \ref{fig2}(a) has the same stoichiometry as the middle-sized nanotube of Fig. \ref{fig1}.
That $C_{4h}$ \BNE hemispherically-capped nanotube is related to this cage
by the insertion of twelve defects: six squares and six octagons.  Using the triple-$\zeta$ basis the energetic
cost is 0.14 eV per defect and using the double-$\zeta$ basis the cost is 0.12 eV per defect, which is
more than an order of magnitude less than the cost of any additional defect in a carbon fullerene.
The calculations indicate that these cages are energetically more stable than the \BNF cage. The trend is
similar to that observed in case of C$_{60}$ where higher fullerenes have larger binding energy than the 
C$_{60}$\cite{DBMMW91}.

\begin{figure}
\epsfig{file=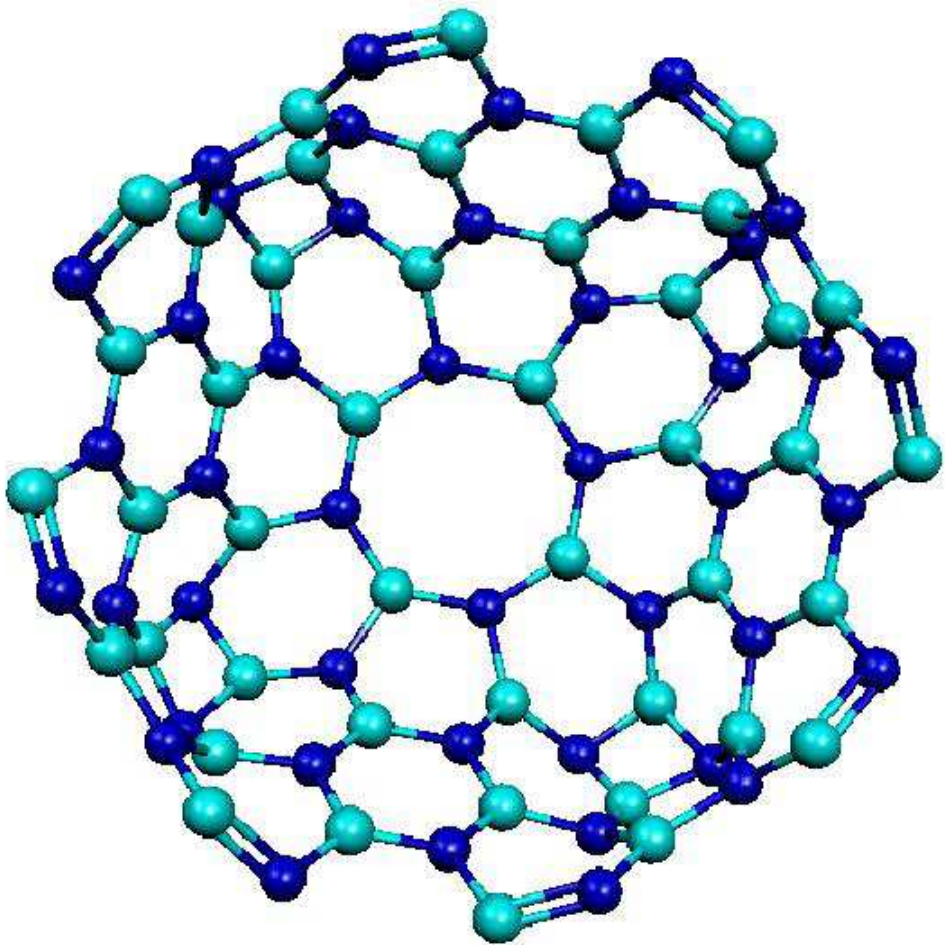,width=\linewidth,clip=true}
\caption{\label{cap:fig} (Color online) The possible round cap of the (8,8) nanotube based on the \BNR cage.}
\end{figure}

C$_{60}$ does not cap the most important carbon nanotubes.  Probably the metallic (10,10) nanotube is
most important \cite{Smalley}.  It can be capped by halves of
round, but not perfectly round like C$_{60}$, icosahedral C$_{240}$ \cite{Harrison}.  \BNF can be enlarged
by adding hexagons.  Another such round cage of octahedral symmetry is \BNR.  The two optimized round 
cages of this study are shown in Fig. \ref{fig3}.  The Table shows that \BNR is the more stable than the \BNF.
Its halves (Cf. Fig. \ref{cap:fig}) would cap a round (8,8) BN nanotube.  It is clearly different, from \BNF$\!$, 
in that while being
mostly round, its twelve squares stick out significantly, like the detonators of a sea mine.  Furthermore,
\BNR is not substantially more stable than the other cages of Table I, despite the fact that it is at least
three times as large.  Thus we think it unlikely that hemispherical caps for larger than BN (4,4)
nanotubes, if they exist at all, will have isolated squares, but no really attractive possibilities come
to mind.  For example, six square-octagon-square units could be placed on the two-fold axes of the
octahedral group, but the Euler overclosure would have to be balanced by two extra octagons inserted
elsewhere to form a hemisphere.  Of course, BN nanotubes could be capped by hemispheres of carbon fullerenes.

          To summarize,  all electron analytic density functional based calculations have been performed on
selected boron-nitride cage structures including the recently synthesized \BNF cage. The energetics
of the BN tubular structures with the end caps based on this  cage 
points to possible existence of round capped BN nanotubes. Amongst the BN cages that satisfy 
the six isolated square rule the higher \BNE and \BNT cages 
are found to be energetically more stable than the smallest \BNF consistent with earlier calculations
\cite{SSL95}, but apparently they are not more
abundant experimentally \cite{Oku03}.  Additional topological defects in the most favorable BN cages seem
more energetically favorable than defects in the fullerenes.  Probably round caps on BN nanotubes with
circular cross section are more likely than flat caps on triangular carbon  nanotubes.

        The Office of Naval Research, directly and through the Naval Research Laboratory, and the 
Department of Defense's   High Performance Computing Modernization Program, through the Common High 
Performance Computing Software Support Initiative Project MBD-5, supported this work.

\end{document}